# Frequency Domain Decomposition Translation for Enhanced Medical Image Translation Using GANs

Zhuhui Wang, Jianwei Zuo, Xuliang Deng, Jiajia Luo

*Abstract*—**Medical Image-to-image translation is a key task in computer vision and generative artificial intelligence, and it is highly applicable to medical image analysis. GAN-based methods are the mainstream image translation methods, but they often ignore the variation and distribution of images in the frequency domain, or only take simple measures to align high-frequency information, which can lead to distortion and low quality of the generated images. To solve these problems, we propose a novel method called frequency domain decomposition translation (FDDT). This method decomposes the original image into a high-frequency component and a low-frequency component, with the high-frequency component containing the details and identity information, and the low-frequency component containing the style information. Next, the high-frequency and low-frequency components of the transformed image are aligned with the transformed results of the high-frequency and low-frequency components of the original image in the same frequency band in the spatial domain, thus preserving the identity information of the image while destroying as little stylistic information of the image as possible. We conduct extensive experiments on MRI images and natural images with FDDT and several mainstream baseline models, and we use four evaluation metrics to assess the quality of the generated images. Compared with the baseline models, optimally, FDDT can reduce Fréchet inception distance by up to 24.4%, structural similarity by up to 4.4%, peak signal-to-noise ratio by up to 5.8%, and mean squared error by up to 31%. Compared with the previous method, optimally, FDDT can reduce Fréchet inception distance by up to 23.7%, structural similarity by up to 1.8%, peak signal-to-noise ratio by up to 6.8%, and mean squared error by up to 31.6%. The results indicate that our method improves the quality of generated images without increasing the computational effort of the baseline model inference.**

## I. INTRODUCTION

MEDICAL image-to-image translation has witnessed significant advancements in recent years and played crucial roles in various clinical applications. It has become an important technique for improving clinical workflows and enabling new diagnosis and treatment planning methods. The ability to synthesize realistic medical images in different modalities from limited input data has significant potential to increase accessibility and reduce the costs of medical imaging [1]. This task involves mapping images from one modality or domain to another, enabling valuable insights and facilitating diagnosis, treatment, and research. For example, magnetic resonance imaging (MRI) and computed tomography (CT) scanners are expensive and not widely available globally. By translating routine clinical 2D ultrasound images into MRI or CT equivalents, quality diagnostic information could be made accessible in low-resource settings [2]. Another application of image translation is virtual enhancement, such as super-resolution and denoising, to improve image quality [3].

MRI is a versatile diagnostic imaging technique available in multiple modalities that provide complementary information about anatomical structures and pathology. Common MRI sequences include T1-weighted, T2-weighted, diffusion-weighted, and contrast-enhanced imaging, each utilizing different tissue properties to generate image contrast [4]. The ability to synthesize realistic MRI scans across modalities from limited input data through cross-translation techniques has important clinical utility. For example, the synthesis of T2 MRI from T1 scans can help radiologists identify tumors, while synthesized contrast-enhanced images can facilitate vascular assessment when contrast agents cannot be used [5]. Cross-translation can also generate multi-parametric MRI to improve diagnosis without prolonging scan time [6]. Furthermore,

This work was supported by National Key R&D Program of China (2022YFC2402103), National Natural Science Foundation of China (31870942), Peking University Clinical Medicine Plus X-Young Scholars Project PKU2020LCXQ017 and PKU2021LCXQ028, and PKU-Baidu Fund 2020BD039. (Corresponding author: Jiajia Luo)

Zhuhui Wang is with the Institute of Medical Technology, Peking University Health Science Center, Beijing 100191, China, also with the Biomedical Engineering Department, Peking University, Beijing 100191, and also with the School of Electrical and Electronic Engineering, Nanyang Technological University, Singapore, 639798, Singapore (e-mail: wang1772@e.ntu.edu.sg).

Jianwei Zuo and Jiajia Luo are with the Institute of Medical Technology, Peking University Health Science Center, Beijing 100191, China, also with the Biomedical Engineering Department, Peking University, Beijing 100191 (e-mails: zuojianwei@stu.pku.edu.cn, jiajia.luo@pku.edu.cn).

Xuliang Deng is with National Engineering Research Center of Oral Biomaterials and Digital Medical Devices, NMPA Key Laboratory for Dental Materials, Beijing Laboratory of Biomedical Materials, Peking University School and Hospital of Stomatology, 100081 P. R. China, also with Department of Geriatric Dentistry, Peking University School and Hospital of Stomatology, Beijing, 100081 P. R. China, and also with Biomedical Engineering Department, Peking University, Beijing, 100191 P. R. China (email: kqdengxuliang@bjmu.edu.cn)

synthesized pseudo-healthy MRI scans may better highlight abnormalities for radiologists [7]. In summary, cross-modal synthesis in MRI has the potential to reduce scan times, improve diagnostic accuracy, and increase the utility of MRI in clinical workflows.

Two main architectures are typically used for medical image-to-image translation: single-input single-output (SISO) and dual-input single-output (DISO). The DISO architecture takes both style and content images as inputs, allowing more flexible control over the translation process. The SISO architecture takes a single input image and generates the corresponding output image, which provides simplicity and ease of implementation and reduces the need for clinical data; in addition, the single input ensures the uniqueness of the output image, which is important in medical image translation tasks.

In recent years, generative adversarial networks (GANs) [8] have emerged as a powerful framework for the image-to-image translation task. GAN-based approaches have shown remarkable success in generating realistic and high-quality images by learning the mapping between different visual domains. One notable work in this area is Pix2Pix [9], proposed by Isola et al., which introduced a conditional GAN architecture that learns the mapping between input images and their corresponding output images. Since then, numerous variants and improvements have been proposed to enhance the performance of GANs in image-to-image translation tasks. For instance, Zhu et al. presented Cycle-Consistent Adversarial Networks (CycleGAN) [10], which incorporates cycle consistency loss to enable unpaired image translation, eliminating the need for paired training data. Similarly, StarGAN [11] introduced a unified GAN model capable of performing multi-domain image translation. Unsupervised image-to-image translation (UNIT) [12], developed by Liu et al., explored unsupervised image translation by learning a shared latent space between domains. In addition, SPADE [13], created by Park et al., introduced a spatially adaptive normalization technique to improve semantic consistency in generated images.

The above works [9-13] lay the basic SISO framework for image translation models based on GANs and also make some optimizations in the spatial domain. To improve the effectiveness of the image translation model, researchers started to constrain the model in the frequency domain. Several works have focused on leveraging the frequency domain properties to improve the quality and robustness of the translation results. Zhang et al. [14] proposed the wavelet knowledge distillation method. For the generator to better generate high-frequency components, this method first decomposes the image into different frequency band representations using the discrete wavelet transform and then performs knowledge distillation on only the high-frequency components. A large GAN is used as a teacher model, and knowledge is transferred from the teacher model to the student model using a student model based on a small Boltzmann machine. The Boltzmann machine is based on a multi-scale representation of the wavelet transform to capture the frequency domain features of an image. A study by Jiang et al. [15] proposed a loss function called focal frequency loss, which first applies a discrete Fourier transform to the input and generated images to transform them to the frequency domain. Then, the loss function is calculated in the frequency domain, emphasizing the most critical frequency components for the reconstruction and synthesis tasks by focusing on details in a specific frequency range and reducing the weight of relatively unimportant frequency components. This guides the model to better learn and retain the important frequency information of the image. Gao [16] found that in cyclic consistency loss, the neural network hides some of the detail information, so features are lost. To recover these features, the high-frequency signal is passed to the decoder, which recovers the high-frequency information to supplement the claimed features through wavelet decomposition. These studies [14-16] show that using frequency domain information can improve the quality of model-generated images, but they suffer from excessive computational overhead and complex model structure. In response to these problems, Cai et al. proposed a new image generation framework based on the frequency domain: FDIT [17]. FDIT uses the frequency domain consistency loss function to constrain the high-frequency information of the translated image to maintain the identity information of the source domain image. The effect of FDIT is verified on multiple datasets since it produces more realistic translated images and offers better identity preservation than the original spatial domain UNIT framework. Although FDIT achieves better frequency domain constraints, there are still some problems with FDIT:

(i) The frequency domain components of the source and generated images are not equal, so the use of L1 loss to force the alignment of the frequency domain components of both will cause distortion in the generated images, which is not in line with the high accuracy requirements of medical image translation tasks.

(ii) FDIT does not constrain the translation process of low-frequency components, which makes the frequency domain constraint incomplete.

(iii) FDIT is based on the DISO architecture, which requires two types of input data, source and reference, thereby increasing the acquisition cost of clinical samples. In addition, the methodology of FDIT is poorly adapted to the SISO framework, while the generated images are susceptible to bias due to reference inputs.

To solve these problems, we propose a novel objective function for SISO frequency-domain image translation inspired by the Geometry-Consistent GAN [18] proposed by Fu et al. Our function is called frequency-domain decomposed translation (FDDT), which treats the high-frequency components of the image as detail or identity information and the low-frequency components as style information. We perform a discrete Fourier transform on the spatial domain image, apply a Gaussian filter to the image in the frequency domain, and subsequently apply an inverse discrete Fourier transform on the filtered image to transform it from the frequency domain back to the spatial domain. We decompose the source-domain image and the generator-generated image into high-frequency components and low-frequency components; subsequently, the high-frequency components and low-frequency components of the source-domain image are input to the same generator again to obtain the translated high-frequency components and the translated low-frequency components. We constrain the consistency between the frequency-domain components of the translated source-domain image and the frequency-domain components of the same frequency band of the generator-generated image to achieve the constraint on the aggregated frequency domain information of the image translation process. Our proposed FDDT comprehensively focuses on both high-frequency and low-frequency information without increasing the size of the model, and the increase in training overhead is relatively small.

Our proposed FDDT can be flexibly combined with the mainstream image-to-image framework without increasing the inference cost of the model. Extensive experiments show that our proposed method can effectively improve the quality of the generated images based on the baseline model. Our main contributions are as follows:

(i) We propose a novel image-to-image objective function, called FDDT, based on the consistency of frequency domain decomposition. FDDT can better reconstruct the frequency domain information of the generated images and enhance the image generation effect. We conduct experiments on the BraTS2021 dataset [19-23] using both paired and unpaired data, and we use CycleGAN, Pix2Pix, and UNIT as the baseline models. The quality of the generated images is improved by adding FDDT, and FDDT outperforms the previous FDIT method in the spatial domain. Compared with CycleGAN, Pix2Pix, and UNIT, the four metrics of Fréchet inception distance (FID) [24], peak signal-to-noise ratio (PSNR) [25], structural similarity (SSIM) [26], and mean squared error (MSE) are all improved by adding FDDT to the model.

(ii) We split the FDDT into a high-frequency part and a low-frequency part, and we add these parts to the UNIT model separately for ablation experiments. The experiments show that constraining the high-frequency information or the low-frequency information alone can improve the quality of image generation based on the baseline. Meanwhile, we find that compared with the previous FDIT method, which only focuses on high-frequency information, the low-frequency information has a greater impact on the overall image quality, and the low-frequency information contains most of the information in the image, so constraining the low-frequency can obtain better image quality than constraining only the high-frequency information.

(iii) We also conduct experiments on the natural image dataset, and the experiments show that the nonlinear FDDT is equally effective in natural images. On the zebra2horse dataset [10], the generation quality of the model is significantly improved compared with the baseline model after adding FDDT, which proves the generalizability of our method. We also investigate the role of the nonlinear module and its influence on the FDDT generation effect. The results show that on natural images, with the increase in the nonlinear operations, the quality of the image generation increases and then tends to stabilize.

## II. METHOD

### A. Background

Generative adversarial network is a very popular image generation method. GAN consists of two parts: generator and discriminator. The goal of the generator is to generate fake samples to deceive the discriminator, and the goal of the discriminator is to determine as accurately as possible whether the samples come from real data or the generator. Generator and discriminator against each other, continuous optimization, and ultimately can be learned to generate the distribution of real data distribution. Our proposed method for image-to-image translation is based on a GAN framework, specifically utilizing the SISO architecture. The SISO framework has shown effectiveness and simplicity in various medical image translation tasks [27, 28], making it suitable for our objective. We define the spatial domain distribution and frequency domain distribution as the following two equations:

$$\mathcal{X} = \mathbb{R}^{H \times W \times C}, \tag{1}$$

$$\mathcal{F} = \mathbb{C}^{H \times W \times C}, \tag{2}$$

Equation (1) defines the spatial domain distribution and Equation (2) defines the frequency domain distribution, where $\mathbb{R}$ represents the datasets of real numbers, $\mathbb{C}$ represents the datasets of complex numbers. $H$, $W$, and $C$ represent the three dimensions of image data: height, width, and channel, respectively.

In recent years, image-to-image translation has gained significant attention in computer vision research. This task aims to learn the mapping between different visual domains, enabling the transformation of images from one domain to another while preserving their semantic content. Several GAN-based approaches have been proposed to tackle this problem, among which three commonly used models are CycleGAN, Pix2Pix, and UNIT. We use these three common models as our baseline models.

CycleGAN is a popular image-to-image translation model that leverages cycle consistency to learn the mapping between two domains without paired training data. It introduces a cycle-consistency loss that encourages the reconstructed image to be close to the original image after going through both forward and backward mappings. This Cycle-Consistent constraint allows CycleGAN to learn meaningful mappings even when direct supervision is unavailable. The formula for the Cycle-Consistent constraint is as follows:

$$\mathcal{L}_{\text{cycle}} = \mathbb{E}_{x_1 \sim \mathcal{X}_1}[\| G_2(G_1(x_1)) - x_1 \|_1] + \mathbb{E}_{x_2 \sim \mathcal{X}_2}[\| G_1(G_2(x_2)) - x_2 \|_1], \tag{3}$$

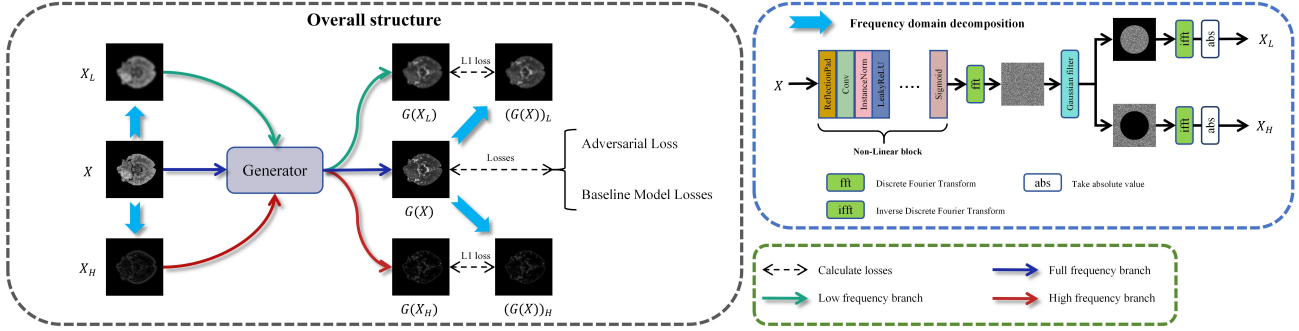

Fig. 1. FDDT structure diagram. The figure mainly shows the overall framework of the FDDT and the calculation of the frequency domain decomposition loss function of the generator. The Non-Linear block of Frequency domain decomposition is based on the convolution module shown in the figure and ends with the sigmoid function. For the BraTS2021 dataset, due to the weak non-linearity of the mapping, it is possible to leave out the nonlinear module or use only the nonlinear module with fewer layers.

where $x_1$ and $x_2$ are images in the source domain $\mathcal{X}_1$ and target domain $\mathcal{X}_2$, respectively; and $G_1$ and $G_2$ are the generators corresponding to mapping $x_1 \rightarrow x_2$ and mapping $x_2 \rightarrow x_1$, respectively.

UNIT leverages the concept of shared latent space to learn the underlying mapping between different domains in an unsupervised manner. UNIT consists of two encoders and two decoders, each dedicated to one domain. The encoders map images from their respective domains to a shared latent space, while the decoders reconstruct the images from the shared latent space. By aligning the latent spaces of the two domains, UNIT enables bidirectional image translation without the need for paired data. It has been successfully applied to various tasks, including style transfer, domain adaptation, and image synthesis. In addition to the basic generative adversarial loss, UNIT's objective function mainly contains variational autoencoder (VAE) loss [29] and Cycle-Consistent constraint, where the VAE loss is as follows:

$$\mathcal{L}_{VAE_1}(E_1, G_1) = \lambda_1 KL\big(q_1(z_1 \mid x_1)||p_\eta(z)\big) - \lambda_2 \mathbb{E}_{z_1 \sim q_1(z_1|x_1)}\big[\log p_{G_1}(x_1 \mid z_1)\big], \tag{4}$$

$$\mathcal{L}_{VAE_2}(E_2, G_2) = \lambda_1 KL\big(q_2(z_2 \mid x_2)||p_\eta(z)\big) - \lambda_2 \mathbb{E}_{z_1 \sim q_2(z_2|x_2)}\big[\log p_{G_2}(x_2 \mid z_2)\big], \tag{5}$$

where $E$ is the encoder that encodes the input vector to produce the intermediate state vector $z$, $KL(\cdot)$ is the operator to compute the $KL$ scatter, and $\lambda_1$ and $\lambda_2$ are weighting factors.

To verify the generalizability of our proposed method, we use both the unpaired image-to-image model and the classical paired image-to-image model using Pix2Pix. Pix2Pix is another widely used image-to-image translation model that learns the mapping from an input image and its corresponding output image using a conditional GAN. It employs a paired training dataset where each input image is paired with its corresponding output image. In addition to the basic adversarial generation loss, the objective function of the Pix2Pix model and adversarial generation loss are shown as follows:

$$\mathcal{L}_{pix}(G_1, G_2) = \mathbb{E}_{x_1 \sim \mathcal{X}_1, x_2 \sim \mathcal{X}_2}[\| x_2 - G_1(x_1) \|_1] + \mathbb{E}_{x_1 \sim \mathcal{X}_1, x_2 \sim \mathcal{X}_2}[\| x_1 - G_2(x_2) \|_1] \tag{6}$$

$$\mathcal{L}_{GAN} = \mathbb{E}_{x_2 \sim \mathcal{X}_2}[\log D_2(x_2)] + \mathbb{E}_{x_1 \sim \mathcal{X}_1}[\log(1 - D_2((G_1(x_1)))) + \mathbb{E}_{x_1 \sim \mathcal{X}_1}[\log D_1(x_1)] + \mathbb{E}_{x_2 \sim \mathcal{X}_2}[\log(1 - D_1(G_2(x_2)))], \tag{7}$$

where $D_1$ and $D_2$ are discriminators with $\mathcal{X}_1$ and $\mathcal{X}_2$ distributions.

In recent years, there have been some works in the field of image translation constraining the mapping and learning of the model in the frequency domain. Proposed by Cai et al., FDIT is a generative framework based on the decomposition of information in the frequency domain of the image. To provide effective image translation, FDIT innovatively focuses on the identity characteristics of the high-frequency information of the image and the low-frequency information of the squiggle characteristics of the image. This approach improves the quality of the generated image and increases the identity information retained in the generated image by preserving the high-frequency information before and after the generation of the image, as well as preserving the coupling between the high- and low-frequency information. Using the FDIT constraints in the spatial domain alone has a similar effect as the full FDIT constraints, and to save computational overhead, we compute the loss function only in the spatial domain. The constraints of FDIT have two main atmospheres: the reconstruction and the translation. The constraint function of FDIT in pixel space is shown as follows:

$$\mathcal{L}_{rec,pix}(E, G) = \mathbb{E}_{x \sim \mathcal{X}}[\| x_L - (G(E(x)))_L \|_1 + \| x_H - (G(E(x)))_H \|_1] \tag{8}$$

$$\mathcal{L}_{trans,pix}(E, G) = \mathbb{E}_{x \sim \mathcal{X}}[\| x_H^{source} - (G(z_c^{source}, z_s^{ref}))_H \|_1] \tag{9}$$

where $x_H^{source}$ is the high-frequency component of the input image in the source domain, $z_c^{source}$ is the content component of the hidden space vector generated by the source domain input image after the encoder $E$, and $z_s^{ref}$ is the style component of the hidden space vector of the target domain image.

### B. Frequency Domain Decomposition Translation

Our novel image translation framework, called FDDT, is shown in Fig. 1. Next, we will introduce the pixel domain training objective function. Our objective function improves the performance of the model without increasing the inference cost. We also discuss the problem that the original FDDT constraints lack the ability to fit nonlinear mappings, especially when the data are natural images, so we propose the nonlinear module to solve this problem.

### C. Model Structure

#### 1) Generator Network

In this study, CycleGAN, Pix2pix, and UNIT are used as baseline models, and the specific structure of the generator varies. For CycleGAN and Pix2pix, the structure of the generator is encoder – decoder structure and ResNet [30] structure, respectively. The first part of the generator contains the encoding path, which is used to extract the feature representation of the input image. It mainly consists of Conv-InstanceNorm [31] rectified linear unit (ReLU), which doubles the number of feature channels at each downsampling, thus increasing the ability of the network to extract features. The middle part contains five residual blocks; each residual block consists of two layers of convolution, each layer of convolution is followed by the InstanceNorm layer, and each residual block is ended by the ReLU activation function, which introduces a shortcut connection to add the input directly to the output of the second layer of convolution. As a result, the convolutional network is able to perform deeper feature extraction and modeling, and the data tensor passes through the residual blocks with unchanged shape. The third part of the generator is an up-sampling decoding path symmetric with the encoding path, which gradually restores the spatial resolution. The last layer uses the Tanh activation function, which maps the output to $[-1, 1]$ as a generated image. Through the encoding – decoding structure, the source domain image can be inputted and converted to the target domain image. In addition, there is no pooling layer in the generator, and downsampling is done using stepwise convolution, which reduces information loss. We use instance normalization to stabilize the GAN training.

For the UNIT model, we use a generator structure called VAE-GAN containing an encoder and a decoder. The encoder contains multi-layer convolutional blocks that are used to extract the input image features. Each layer of convolution is followed by InstanceNorm-ReLU, which employs downsampling convolution to obtain higher-level semantic features. After downsampling, multiple residual blocks are connected in series to continue extracting features, and the final output is a low-dimensional vector representation of the image features as a hidden variable. The decoder is symmetric with the encoder structure and maps the hidden variables output from the encoder back to the image space after residual blocks and up-sampling. The last layer uses the Tanh activation function to output the reconstructed image. A PatchGAN discriminator [12] is used in both the encoder and decoder to force the hidden variables and the generated sample distribution to match the target distribution.

#### 2) Discriminator Network

The discriminator structure used in this paper is called PatchGAN discriminator. For CycleGAN and Pix2Pix, the discriminator is a five-layer CNN network with convolutional kernel sizes of $4 \times 4$. The first convolutional layer is Conv-LeakyReLU, the next three convolutional layers are Conv-InstanceNrom-LeakyReLU, and the last convolutional layer has only convolutional operations with 64, 128, 256, 512 filters. Finally, we carry out average pooling operations to obtain the feature maps used to compute the discriminative loss. For the UNIT model, we construct a multi-scale discriminator, perform average pooling on the input image, downsample it to three different scales, and input each scale into three separate PatchGAN discriminators to compute the discriminative loss for each scale, which makes the judgment more comprehensive and accurate.

#### 3) Non-Linear Block

We note that the FDDT method suffers from a more pronounced lack of nonlinearity while constraining the frequency domain information transformation. We assume that the image-to-image transformation can be linear or nonlinear, and for nonlinear transforms, the FDDT can be fit it better. The discrete Fourier transform is known to be a linear transform, and Gaussian filtering is also linear [32]. The objective function of the FDDT has the following properties:

$$\begin{cases} G(\mathcal{F}_H(X)) = \mathcal{F}_H(G(X)) \\ G(\mathcal{F}_L(X)) = \mathcal{F}_L(G(X)) \\ X = \mathcal{F}_H(X) + \mathcal{F}_L(X) \end{cases} \quad (10)$$

In the above equation, $X$ is the image input to the generator, and $\mathcal{F}_H$ and $\mathcal{F}_L$ are the discrete Fourier high-frequency filter and discrete Fourier low-frequency filter, respectively. From the above equation, we can derive the following equation:

$$G(\mathcal{F}_H(X) + \mathcal{F}_L(X)) = G(\mathcal{F}_H(X)) + G(\mathcal{F}_L(X)). \quad (11)$$

To reduce the constraints of FDDT on the superposition of generator functions, we use a smaller CNN network as a precursor module for fitting the nonlinear mapping, as shown in Fig. 1. The basic convolution module consists of ReflectionPad -Conv-InstanceNorm-LeakyReLU, the convolution kernel size is 3, the number of filters is twice the number of input image channels, and the feature image is output using the Sigmoid activation function.

### D. Loss Function

This study focuses on improving the training part of the generator of GANs. We first define the spatial domain $\mathcal{X} = \mathbb{R}^{H \times W \times 1}$ and generators G given a source domain image $X \in \mathcal{X}$. The frequency domain decomposition loss function of the generator can be written in the following form:

$$\mathcal{L}_{freq}(G) = \mathbb{E}_{X \sim \mathcal{X}} \left[ \parallel G(X_L) - (G(X))_L \parallel_1 \right] + \\ \mathbb{E}_{X \sim \mathcal{X}} \left[ \parallel G(X_H) - (G(X))_H \parallel_1 \right], \quad (12)$$

where $X_L$ and $X_H$ are the low-frequency component and the high-frequency component of the source domain image $X$, respectively. The frequency domain component is arrival from the discrete Fourier transform with Gaussian filtering. The

Gaussian filter formula we use is shown as follows:

$$H_L(u,v) = \frac{1}{2\pi\sigma^2} e^{-\frac{u^2+v^2}{2\sigma^2}}, \qquad (13)$$

$$H_H(u,v) = 1 - H_L(u,v), \qquad (14)$$

where $H_L(u,v)$ and $H_H(u,v)$ are low-pass and high-pass filters, respectively, $\sigma$ is the standard deviation of the filter, which we set to 20, and $(u,v)$ are the coordinates in the spectrogram. We first perform a discrete Fourier transform on a spatial domain image $X$ of shape $H \times W \times 1$ and map it to the frequency domain:

$$\mathcal{F}(X)(u,v) = \frac{1}{HW} \sum_{h=0}^{H-1} \sum_{w=0}^{W-1} e^{-2\pi i \frac{hu}{H}} e^{-2\pi i \frac{wv}{W}} \cdot N(X(h,w)), \qquad (15)$$

where $N(*)$ is non-linear mapping used to increase the model's ability to fit non-linear mappings. We use shallow neural networks with ReLU and sigmoid activation functions as nonlinear modules.

We then use Gaussian filter $X$ in the frequency domain, and the following are the frequency domain images after low-pass filtering and high-pass filtering, respectively:

$$\mathcal{F}_L(X)(u,v) = \mathcal{F}(X)(u,v) \cdot H_L(u,v) \qquad (16)$$

$$\mathcal{F}_H(X)(u,v) = \mathcal{F}(X)(u,v) \cdot H_H(u,v) \qquad (17)$$

Finally, we map the image from the frequency domain back to the spatial domain by inverse discrete Fourier transform. To eliminate abnormal pixel values due to truncation effects and computer errors, as well as to add some nonlinear components, we take the absolute value of the resulting image:

$$X_L(h,w) = \left| \frac{1}{HW} \sum_{u=0}^{H-1} \sum_{v=0}^{W-1} \mathcal{F}_L(X)(u,v) e^{i2\pi\left(\frac{uh}{H} + \frac{vw}{W}\right)} \right|, \qquad (18)$$

$$X_H(h,w) = \left| \frac{1}{HW} \sum_{u=0}^{H-1} \sum_{v=0}^{W-1} \mathcal{F}_H(X)(u,v) e^{i2\pi\left(\frac{uh}{H} + \frac{vw}{W}\right)} \right|. \qquad (19)$$

The total loss function of the model generator is shown as follows:

$$\mathcal{L}_{total}(G) = \lambda_1 \mathcal{L}_{baseline}(G) + \lambda_2 \mathcal{L}_{freq}(G), \qquad (20)$$

where $\mathcal{L}_{baseline}(G)$ is the loss function of the generator of the baseline model. In this study, $\lambda_1 = \lambda_2 = 1$.

### E. Datasets

In this study, we conduct experiments using the BraTS2021 dataset, a widely recognized benchmark dataset in medical imaging. The BraTS2021 dataset consists of brain MRI scans obtained with five different modalities, namely T1, T1c, T2, Flair, and segmentation masks. For our study, we specifically utilize the T1, T1c, Flair and T2 modalities, as they are known to provide valuable information for brain tumor segmentation and classification tasks. The images in the BraTS2021 dataset have a standardized size of pixels, and they were acquired using MRI. T2-weighted imaging is sensitive to edema and tumor infiltration, and Flair imaging enhances the visibility of lesions by suppressing the cerebrospinal fluid signal. T1-weighted images can clearly display the adipose tissue and glandular structure of the human body, have good contrast for soft tissues, and have high resolution. T1c mode uses contrast agent to enhance the contrast based on T1, which can display the lesion more clearly. Mainly used for the diagnosis of brain tumors and other lesions.

Translation between different modalities of MRI holds significant clinical relevance. For example, transforming T2 images to Flair images can enhance the visualization of tumor-associated edema, which is crucial for diagnosing and monitoring brain tumors. Conversely, translating Flair images to T2 images can help accurately identify tumor boundaries and characterize their extent. By bridging the gap between modalities, our proposed method aims to contribute to more comprehensive and accurate clinical assessments, thereby assisting medical professionals in making informed decisions for brain tumor diagnosis and treatment planning. Converting T1 modality MRI images to T1c modality images can simulate the T1c enhancement effect in a non-invasive way, avoiding repeated injections of contrast agents to patients. The acquisition of T1 data is relatively easy, and the converted synthetic T1c images can be used to assist Diagnose oncological diseases.

The dataset contains paired images, meaning that each Flair image corresponds to a T2 image, each T1 image corresponds to a T1ce image, allowing for direct comparison and analysis. This paired nature of the dataset enables us to leverage the inherent correlation between different modalities and explore their translation capabilities. We randomly select a subset of the BraTS2021 dataset on which to conduct our experiments. The training set consists of 3,840 pairs of Flair-T2 images and T1-T1ce images, and the test set consists of 960 pairs. This subset provides a sufficiently large and diverse collection of images to effectively train and evaluate our proposed method.

By utilizing the BraTS2021 dataset, selecting appropriate subsets, and considering the clinical significance of different modalities, we aim to validate the effectiveness and applicability of our proposed image translation framework FDDT in enhancing medical image analysis and improving patient care in the context of brain tumor imaging.

In addition to evaluating the performance of FDDT on medical images, we conduct experiments on the well-established zebra2horse dataset to validate its effectiveness in handling natural images. The zebra2horse dataset is a widely used benchmark dataset for image-to-image translation tasks in the field of computer vision. Images in the zebra2horse dataset have a standardized size of $3 \times 256 \times 256$ pixels, ensuring consistency and facilitating direct comparison between different methods. By employing the zebra2horse dataset alongside medical images, we aim to demonstrate the versatility and effectiveness of FDDT across different domains. The inclusion of natural images allows us to evaluate the model's ability to capture and translate complex visual patterns, textures, and structures.

### F. Model Training

We train CycleGAN, UNIT, and Pix2Pix on the BraTS2021 dataset as baseline models, and we train them again for comparison after adding the FDDT objective function and a variant of the FDIT objective function to them. On the

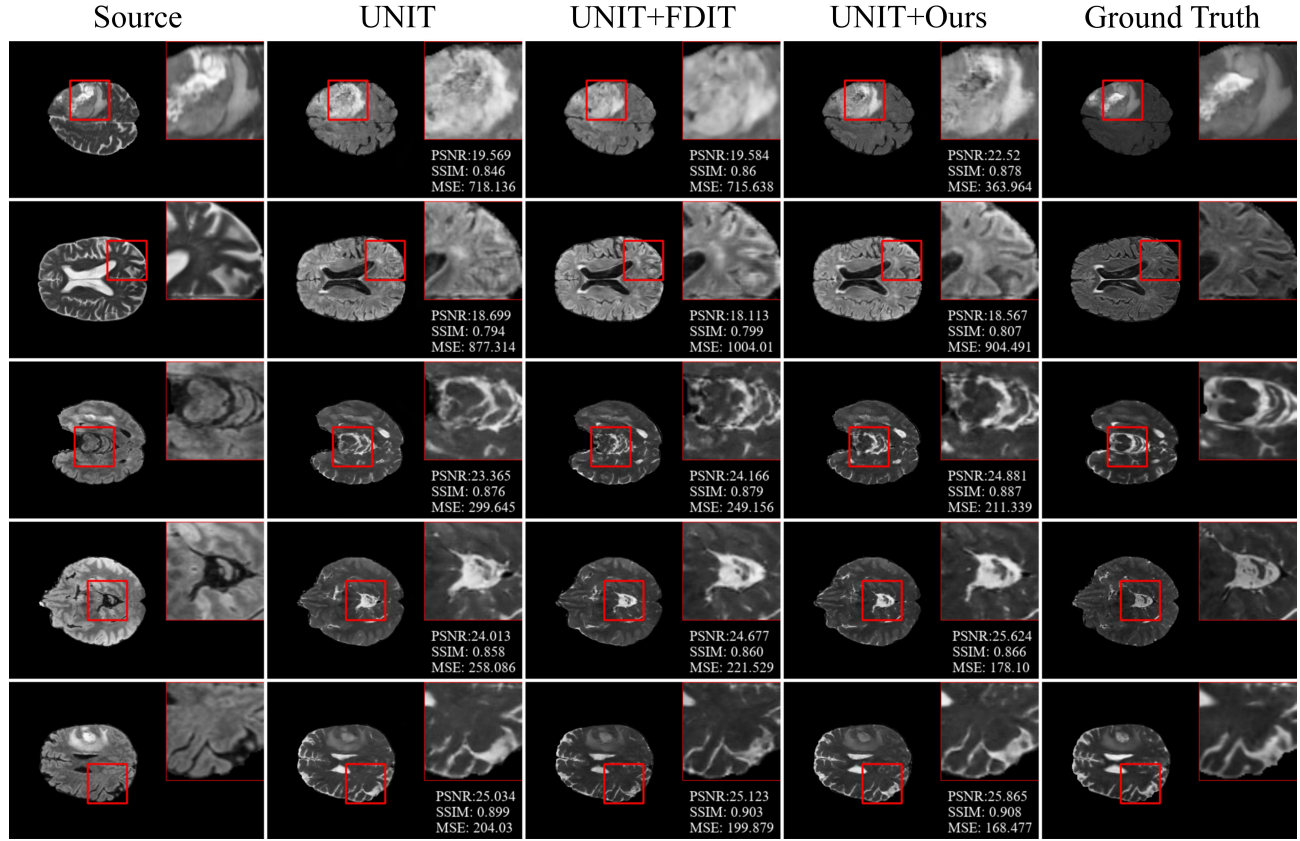

Fig. 2. Comparison of generated image results. Our proposed method can enhance the quality of generated images in the SISO system. Compared with the baseline model (i.e., UNIT) and previous method (i.e., spatial domain FDIT), our method can retain more texture information and detail information, especially in the key areas (e.g., lesions). It can also produce an image that is closer to the real image in terms of style information.

zebra2horse dataset, we train the model with the addition of frequency domain objective functions (FDDT, FDIT) with the addition of different nonlinear modules using only the CycleGAN model as the baseline. In the above training process, we use the Adam optimizer with $\beta_1 = 0.5, \beta_2 = 0.999$. The learning rate is 0.0001, batch size is 32, total number of epochs is 300, and after the 200th epoch, the learning rate linearly and progressively decays to 0. All models are run using Pytorch (v1.12.0) and Python (v3.8.10).

### G. Evaluation Metrics

#### 1) Full-Reference Metrics

For the dataset with aligned data (BraTS2021), we use MSE, PSNR, and SSIM to evaluate the image translation quality. The MSE reflects the error between the real image and the generated image at the pixel level, which is calculated as follows:

$$MSE = \frac{1}{mn} \sum_{i=0}^{m-1} \sum_{j=0}^{n-1} [I(i,j) - K(i,j)]^2, \quad (21)$$

where $I$ is the real image, $K$ is the generated image, $m, n$ are the height and width of the image, respectively, and a smaller MSE value means the two images are more similar.

PSNR reflects the signal quality, and its formula is:

$$PSNR = 10 log_{10} \frac{R^2}{MSE}, \quad (22)$$

where $R$ is the peak pixel value of the image, and MSE is the mean square error. The larger the PSNR, the better the signal-to-noise ratio, and the more similar the generated image is to the real image. PSNR is easy to compute but relies too much on pixel-level error and lacks structural similarity considerations.

SSIM considers the structural information of the image. Based on cognitive theory, the image's BRIGHT, CONTRAST, and STRUCTURE are taken into account to calculate the three similarities [26]. SSIM is in the interval of $[0, 1]$, where a larger value indicates greater similarity between the structures of the two images. The formula is:

$$SSIM(x, y) = \frac{(2\mu_x\mu_y + c_1)(2\sigma_{xy} + c_2)}{(\mu_x^2 + \mu_y^2 + c_1)(\sigma_x^2 + \sigma_y^2 + c_2)}, \quad (23)$$

where x and y are the two images to be compared; $\mu_x$ and $\mu_y$ are the pixel means of x and y, respectively; $\sigma_x$ and $\sigma_y$ are the standard deviations of the pixel values of x and y, respectively; $\sigma_{xy}$ is the covariance of the pixel values of x and y; and $c_1$ and $c_2$ are two very small constants used for stabilization calculation.

#### 2) No-Reference Metrics

For the problem of image translation without strictly paired and aligned data, FID is often used as an evaluation metric for the quality of image translation. FID is based on the Fréchet distance of a Gaussian distribution, and it calculates the generative power of the model by comparing the similarity between the real image distribution and the generated image distribution. The formula is as follows:

$$FID = \| \mu_g - \mu_r \|^2 + Tr\left(\Sigma_g + \Sigma_r - 2\left(\Sigma_g \Sigma_r\right)^{1/2}\right) \quad (24)$$

where $\mu_g$ and $\mu_r$ are the mean vectors of the generated data and the real data, respectively; $Tr$ denotes the trace of the matrix; and $\Sigma_g$ and $\Sigma_r$ are the covariance matrices of the generated data and the real data, respectively. The smaller the value of FID, the more similar the generated image is to the real image. It should be noted that all the parameters in the above equation are obtained in the feature space; that is, the images are mapped to the same feature space in advance using the pre-trained Inception v3 network [33].

Table I
COMPARATIVE EXPERIMENTS ON TRANSITIONS BETWEEN T2 AND FLAIR MODALITIES.
USE BOLD TO INDICATE RELATIVELY OPTIMAL RESULTS.

| Method | | SSIM↑ | PSNR↑ | MSE↓ | FID↓ |
|---|---|---|---|---|---|
| CycleGAN | T2 to Flair | 0.813±0.003 | 22.383±0.142 | 421.133±18.490 | 49.412±3.339 |
| | Flair to T2 | 0.830±0.001 | 22.490±0.025 | 387.523±2.387 | 40.637±1.258 |
| CycleGAN+FDIT | T2 to Flair | 0.837±0.004 | 22.074±0.337 | 471.861±39.488 | 42.735±3.259 |
| | Flair to T2 | 0.839±0.002 | 22.756±0.131 | 362.783±8.167 | 47.073±1.040 |
| CycleGAN+Ours | T2 to Flair | **0.849±0.001** | **22.783±0.222** | **408.273±21.705** | **38.646±1.343** |
| | Flair to T2 | **0.854±0.001** | **23.351±0.055** | **321.293±3.896** | **35.899±0.770** |
| UNIT | T2 to Flair | 0.854±0.001 | 22.700±0.072 | 439.524±7.633 | 35.924±0.333 |
| | Flair to T2 | 0.863±0.001 | 23.660±0.034 | 299.692±1.917 | 23.720±0.801 |
| UNIT+FDIT | T2 to Flair | 0.859±0.001 | 22.940±0.075 | 432.927±8.148 | 30.239±0.596 |
| | Flair to T2 | 0.866±0.001 | 23.857±0.099 | 293.337±6.095 | **21.928±0.496** |
| UNIT+Ours | T2 to Flair | **0.862±0.001** | **22.975±0.035** | **416.393±2.138** | **27.147±0.264** |
| | Flair to T2 | **0.867±0.001** | **23.899±0.134** | **284.998±9.519** | 25.57±1.144 |
| Pix2Pix | T2 to Flair | 0.831±0.004 | 21.967±0.081 | 476.283±11.157 | 44.345±0.649 |
| | Flair to T2 | 0.832±0.002 | 23.0657±0.091 | 338.679±6.925 | 35.442±0.322 |
| Pix2Pix+FDIT | T2 to Flair | 0.824±0.001 | 21.932±0.083 | 483.250±11.015 | **43.511±0.746** |
| | Flair to T2 | 0.836±0.001 | 23.064±0.013 | 342.815±1.410 | 30.739±0.399 |
| Pix2Pix+Ours | T2 to Flair | **0.833±0.002** | **22.434±0.126** | **423.832±16.120** | 43.791±1.685 |
| | Flair to T2 | **0.842±0.003** | **23.265±0.129** | **323.107±9.660** | **29.121±0.387** |

Table II
COMPARATIVE EXPERIMENTS ON TRANSITIONS BETWEEN T1 AND T1CE MODALITIES.
USE BOLD TO INDICATE RELATIVELY OPTIMAL RESULTS.

| Method | | SSIM↑ | PSNR↑ | MSE↓ | FID↓ |
|---|---|---|---|---|---|
| CycleGAN | T1 to T1ce | 0.874±0.007 | 23.759±1.056 | 366.095±112.484 | 38.342±0.625 |
| | T1ce to T1 | 0.888±0.004 | 23.507±0.678 | 387.576±72.420 | 81.245±7.092 |
| CycleGAN+FDIT | T1 to T1ce | 0.873±0.007 | 23.538±0.477 | 393.174±40.342 | 41.875±1.343 |
| | T1ce to T1 | 0.886±0.004 | 23.336±0.527 | 404.340±63.906 | **75.507±4.521** |
| CycleGAN+Ours | T1 to T1ce | **0.882±0.008** | **25.141±0.947** | **269.051±50.723** | **37.631±2.325** |
| | T1ce to T1 | **0.891±0.002** | **23.703±0.109** | **350.494±11.624** | 80.950±2.446 |
| UNIT | T1 to T1ce | 0.869±0.002 | 24.860±0.309 | 361.458±4.542 | 39.426±0.494 |
| | T1ce to T1 | 0.872±0.001 | 23.545±0.041 | 375.899±5.559 | 93.475±0.566 |
| UNIT+FDIT | T1 to T1ce | 0.879±0.001 | 24.915±0.228 | 272.806±14.098 | 39.220±0.453 |
| | T1ce to T1 | 0.874±0.001 | 23.595±0.228 | **361.572±21.426** | 77.820±0.721 |
| UNIT+Ours | T1 to T1ce | **0.888±0.001** | **25.597±0.183** | **249.321±9.905** | **37.944±2.237** |
| | T1ce to T1 | **0.879±0.001** | **23.609±0.199** | 375.263±17.400 | **71.738±2.875** |
| Pix2Pix | T1 to T1ce | 0.837±0.001 | 23.068±0.183 | 438.121±21.282 | **52.996±0.371** |
| | T1ce to T1 | 0.864±0.001 | 23.051±0.091 | 399.065±11.115 | 92.434±1.825 |
| Pix2Pix+FDIT | T1 to T1ce | 0.839±0.002 | 23.496±0.258 | 409.450±18.558 | 59.893±0.480 |
| | T1ce to T1 | 0.864±0.001 | **23.294±0.023** | **386.973±1.375** | 91.912±1.348 |
| Pix2Pix+Ours | T1 to T1ce | **0.847±0.001** | **24.137±0.022** | **319.240±0.941** | 54.930±0.478 |
| | T1ce to T1 | **0.865±0.001** | 23.090±0.180 | 442.351±22.558 | **86.501±0.174** |

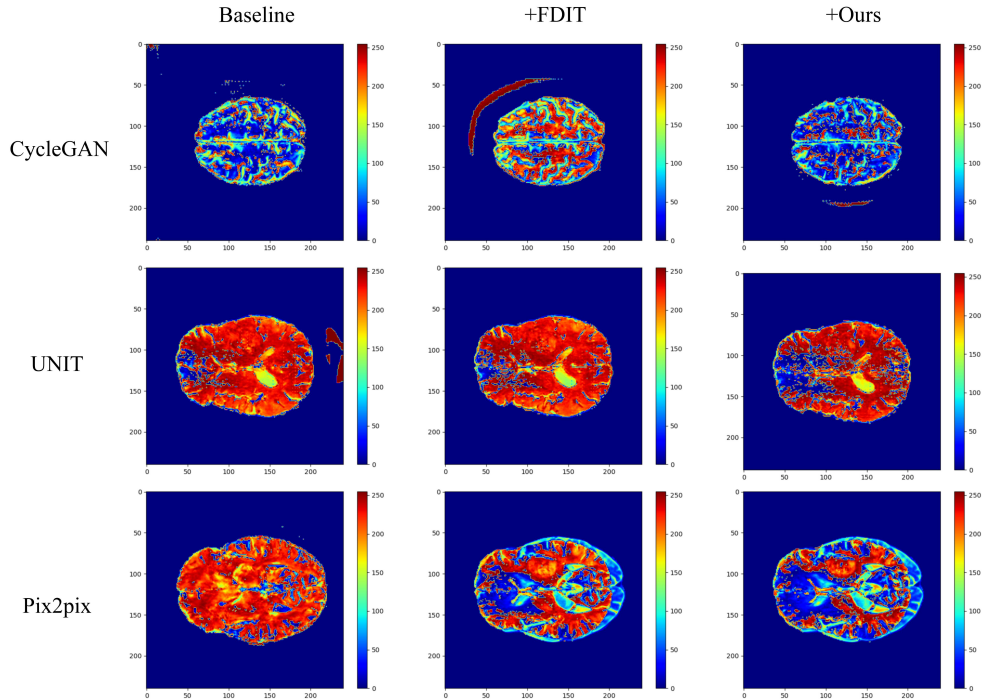

Fig. 3. Absolute error distribution between images generated by baseline models and images generated by baseline+FDDT with ground truth. For unpaired image translation models (CycleGAN, UNIT), the addition of FDDT reduces the absolute error with respect to the ground truth in the entire pixel space, and especially in the low-frequency portion the error can be significantly reduced (low-frequency portion in the center of the brain MRI image). For the paired image translation model (Pix2Pix), the addition of FDDT improves the effect of the model relatively little, and it occasionally increases the error of low-frequency information, but it effectively reduces the error of the high-frequency information, so the edge information and detail information of the generated image are more realistic. Pix2Pix+FDIT uses high-frequency losses in the air domain from FDIT and low-frequency losses from FDDT.

## III. EXPERIMENTS

To verify the effectiveness and feasibility of our method, we conduct experiments on the medical image dataset BraTS2021 and the natural image dataset zebra2horse. In this study, the baseline models used are CycleGAN, Pix2Pix, and UNIT. A large number of experiments show that our proposed FDDT effectively improves the quality of image generation.

### 1) Comparison Experiments

This section presents a comprehensive experimental comparison of our proposed method, which incorporates a new objective function called FDDT, with the baseline model and a previously proposed objective function called FDIT. We use the BraTS2021 dataset for the experiments. To evaluate the performance of the different models, we employ several widely used evaluation metrics, including SSIM, PSNR, FID, and MSE.

First, we compare the results of the baseline models incorporating FDIT and FDDT. The baseline models are popular image-to-image translation approaches without any additional objective functions. Our goal is to investigate the impact of integrating the two proposed objective functions on the overall performance. The results of the experiment are shown in Table I, Table II and Fig. 2.

Our experimental results demonstrate that both FDIT and FDDT contribute to improving the quality of the translated images compared with the baseline models. However, a notable finding is that for each baseline model, the model incorporating FDDT outperforms the model incorporating FDIT in terms of all evaluation metrics. This suggests that FDDT captures more meaningful features in the case of constrained spatial domains only, leading to more accurate and visually appealing image translations.

Furthermore, we observe that the model incorporating FDDT achieves significant improvements compared with the baseline model across most evaluation metrics. The SSIM and PSNR scores indicate a higher level of similarity and fidelity between the translated and ground truth images. The FID score, which measures the distribution divergence, shows that the model incorporating FDDT generates images that are closer in distribution to the real images. In addition, the MSE score demonstrates reduced pixel-wise error in the translated images, indicating better reconstruction accuracy. These results collectively demonstrate the effectiveness of our proposed FDDT objective function in enhancing the performance of the baseline image-to-image translation model. The incorporation of FDDT leads to superior results compared with both the baseline model and the model incorporating FDIT.

From the absolute error heat map (Fig. 3), our model can better reduce the error of frequency information, such as the central location of the brain. At the same time, the high-frequency detail information, such as contours and edcan be well preserved. For the non-paired image translation model, our method can achieve significant results, but for the paired image translation model, our method achieves relatively fewer gains. This is because the paired images themselves already have strong constraints.

Table III
UNIT-BASED FDDT ABLATION EXPERIMENTS.
UNIT+HIGH FREQ AND UNIT+LOW FREQ REPRESENT THE HIGH-FREQUENCY AND LOW-FREQUENCY COMPONENTS ADDED
TO THE FDDT ALONE, RESPECTIVELY.

| Method | SSIM↑ | PSNR↑ | MSE↓ | FID↓ |
|---|---|---|---|---|
| UNIT | 0.854 | 22.700 | 439.524 | 35.924 |
| UNIT+FDDT | 0.862 | 22.975 | 416.393 | 27.147 |
| UNIT+High Freq | 0.856 | 22.828 | 428.104 | 29.645 |
| UNIT+Low Freq | 0.863 | 22.922 | 420.318 | 28.150 |

### 2) Ablation Experiments

In this section, we conduct an ablation study to investigate the individual contributions of the high-frequency and low-frequency components of our proposed FDDT objective function. By analyzing their effects independently, we aim to gain insights into the significance of each component and assess their combined impact on the image-to-image translation performance. To conduct the ablation study, we compare three variants of our model: (1) the baseline model without incorporating FDDT, (2) the model incorporating only the high-frequency component of FDDT, (3) the model incorporating only the low-frequency component of FDDT and (4) the model incorporating both the high-frequency and low-frequency components of FDDT. We use UNIT as the baseline model. The experimental results are presented in Table III.

We can observe that the inclusion of both the high-frequency component and the low-frequency component improves the image generation of the baseline model. Comparing the baseline model with the model using only the low-frequency component, there is a significant improvement in several evaluation metrics. This improvement indicates that the low-frequency component captures and processes low-frequency information separately, thus improving image quality. Furthermore, we observe that the inclusion of the high-frequency component in addition to the low-frequency component further improves the translation performance compared with the baseline model and the model using only the low-frequency component. The inclusion of high-frequency components helps to retain low-frequency detail information, thus making the translation more visually coherent and realistic. In addition, the low-frequency component contributes more to FDDT and improves the model's performance more significantly, whereas the previous FDIT framework directly constrains the translation of the low-frequency component of the image.

### 3) Non-Linear Study on Natural Image Datasets

To verify the extensiveness and generalizability of our proposed FDDT method, we also conduct comparison experiments on the zebra2horse dataset. The experiments show that our method works well on natural images and improves the image generation quality of the baseline model, and the model obtains significant improvement in FID metrics.

We find that the nonlinearity of the nonlinear module in FDDT needs to be increased on natural image datasets, and the model with the nonlinear module is better than without the nonlinear module, but the gain from increasing the number of convolutional layers of the nonlinear module is not significant, and it reaches the elbow point when the number of layers is 1. Since the transformation between natural images is more complex and nonlinear, it is necessary to increase the number of network layers of the nonlinear module to better fit the mapping between source and target domains that is extracted in the feature space.

To identify which operations the nonlinear module implements on the input image, we map the input image to the feature space using the trained single-layer nonlinear module. The results are presented in Fig. 4. We find that the non-downstream retrograde module mainly performs filtering and localization operations on the image, i.e., highlighting the edge high-frequency information and smoothing the low-frequency background information, which facilitates the subsequent frequency-domain decomposition.

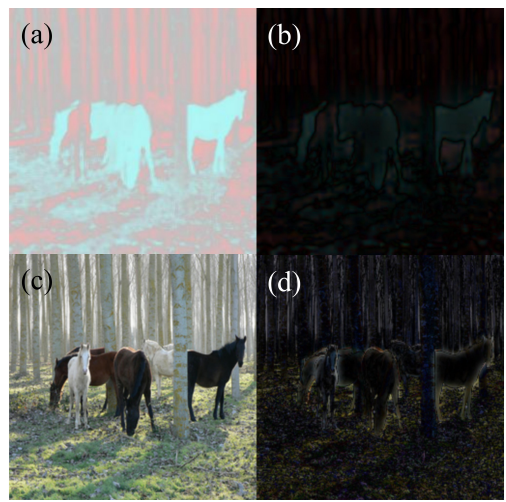

Fig. 4. Source domain image and nonlinearly processed image and their respective high-frequency components. We use CycleGAN and the one-layer CNN network nonlinear module for FDDT training. (a) The feature map obtained when the source domain image is fed into the nonlinear module. (b) The high-frequency component of the feature image. (c) The source domain image. (d) The high-frequency component of the originating image. We can see that extracting high-frequency features in the feature space can filter out more redundant information (e.g., tree trunks, grass).

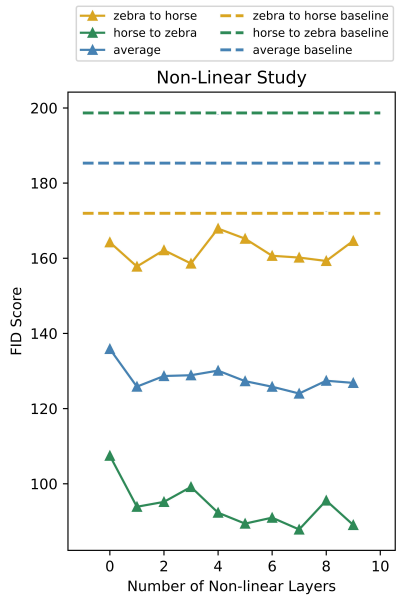

Fig. 5. We find that as the number of convolutional modules increases, the image generation improves and the model reaches saturation when the non-linear module contains one layer of convolutional modules. Adding FDIT constraints results in FID scores that outperform the baseline model.

## IV. DISCUSSION

Multi-modal medical images provide more information than single-mode images, so they can increase the accuracy of clinical diagnosis. However, acquiring multi-modal medical images often imposes additional costs. One solution is to use image translation methods to generate target modalities from source modalities. However, existing medical image translation methods often neglect to properly constrain the frequency domain information, so in this study we aim to develop a new medical image translation method to overcome the above difficulties.

For our proposed FDDT method, we construct an objective function that constrains the consistency of the frequency domain information before and after translation and incorporates a CNN network before the frequency domain transformation to enhance the model's ability to fit nonlinear mappings and extract features in the frequency domain. We take popular image-to-image models (i.e., CycleGAN, UNIT, Pix2Pix) as the baseline models and add the FDDT objective function to each. Compared with the previous frequency domain translation method (FDIT) and the baseline models, our method gives better results on medical images and natural images. We compare the results of images generated by different methods on several quantitative metrics, and FDDT gives optimal results. Our method differs from other frequency domain-constrained methods in two main ways: (i) it inputs the high-frequency component and the low-frequency component of the source domain image into the generator separately to get the translated frequency domain component and align it with the same frequency band component of the generated image; (ii) it incorporates a nonlinear module consisting of a multi-layered CNN network before applying a discrete Fourier transform to the input image. According to Table I, when using CycleGAN and Pix2Pix as the baseline, our method outperforms previous methods in all three full-reference metrics, namely, SSIM, PSNR, and MSE, as well as the FID metric; when using UNIT as the baseline, our method outperforms previous methods in all three full-reference metrics, and it is similar to previous methods in the FID metric.

Previous image generation methods using frequency domain constraints tend to directly transform the image to the frequency domain and then use traditional filtering to extract features or use a loss function to constrain the model directly in the frequency domain, which often requires manually selecting the appropriate feature bands according to the features of different datasets, and at the same time, the redundant information in the image (noise, background, etc.) will be retained in the spectrogram after the frequency domain transformation, which will increase the difficulty of model learning. By contrast, our method maps the input image into the feature space by constructing a multi-layer CNN network and then performs a discrete Fourier transform on the image in the feature space to transform it from the feature space to the frequency domain space. This can adaptively select the feature frequency band and filter out some redundant information. It also enhances the ability of the model to fit nonlinear mapping.

The FDIT method proposed by Cai et al. uses frequency domain decomposition generation in an image translation task by constructing perceptual and reconstruction losses to make the image content of the image style domain merge. There are significant differences between our method and FDIT. (i) Our method is a SISO method, which requires only a single modality as model input, thus reducing the required quantity of data and avoiding errors caused by differences in stylistic inputs. FDDT can be better adapted to mainstream SISO image translation models. (ii) The FDIT method mainly targets unpaired natural images, which do not require high pixel-level accuracy of the generated images. For medical images with strong alignment, we observe that the high-frequency components (content information) of the source and target domains are not the same, so we also input the frequency-domain components of the input images into the model and constrain the generation of frequency-domain features by constructing the frequency-domain decomposition consistency loss. Our approach has some advantages in both quantitative metrics and model fitness.

Currently, we use the traditional discrete Fourier transform (DFT) and single Gaussian kernel function filtering for frequency domain decomposition, but this approach leads to several problems: (i) basis functions used in DFT are sinusoidal, which may not be optimal for representing all image contents [34]; and (ii) DFT has poor spatial localization, making it difficult to pinpoint the spatial origin of frequency components [35]. Although DFT is still widely used in this signal transformation domain, trying other frequency domain transformation methods such as discrete cosine transform (DCT), discrete wavelet transform [35], and contourlet transform [34] can better decompose the frequency domain information. In this work, we take the frequency domain components of the image ground as input and construct the frequency domain decomposition consistency objective function to guide the generator to learn the frequency domain features. However, using only one common generator network may increase the learning burden of the network, and multiple separate neural networks can be used to learn the generation of different frequency bands when sufficient computational resources are available.

## V. CONCLUSION

In this paper, a new SISO image transformation constraint method, FDDT, is proposed, which simultaneously constrains the high-frequency information and low-frequency information before and after image transformation in the spatial domain using network learning. Unlike previous frequency domain constraints, FDDT explicitly constrains both the low-frequency and high-frequency information of the image; FDDT uses network learning to constrain the frequency domain information before and after image transformation, which effectively avoids the inconsistency of the frequency domain information before and after image transformation. Experimental results on a medical image dataset and a natural image dataset show that FDDT effectively improves the quality of image generation. For natural image translation with a strong nonlinear nature, adding a nonlinear module to FDDT can effectively avoid the underfitting phenomenon caused by linear constraints.